\documentclass[conference,10pt]{IEEEtran}
\IEEEoverridecommandlockouts
\AtBeginDocument{%
  \providecommand\BibTeX{{%
    Bib\TeX}}}

\usepackage{graphicx}
\usepackage{textcomp}
\usepackage{xcolor}
\definecolor{dkgreen}{rgb}{0,0.6,0}
\definecolor{gray}{rgb}{0.5,0.5,0.5}
\definecolor{mauve}{rgb}{0.58,0,0.82}
\definecolor{OliveGreen}{rgb}{0,0.6,0}
\definecolor{mahogany}{rgb}{0.75, 0.25, 0.0}
\definecolor{darkmidnightblue}{rgb}{0.0, 0.2, 0.4}
\definecolor{navyblue}{rgb}{0.0, 0.0, 0.5}
\definecolor{apricot}{rgb}{0.98, 0.81, 0.69}
\definecolor{antiquewhite}{rgb}{0.98, 0.92, 0.84}
\definecolor{brickred}{rgb}{0.8, 0.25, 0.33}
\definecolor{bananamania}{rgb}{0.98, 0.91, 0.71}
\definecolor{bisque}{rgb}{1.0, 0.89, 0.77}
\definecolor{champagne}{rgb}{0.97, 0.91, 0.81}
\definecolor{eggshell}{rgb}{0.94, 0.92, 0.84}
\definecolor{darkolivegreen}{rgb}{0.33, 0.42, 0.18}
\definecolor{phthalogreen}{rgb}{0.07, 0.21, 0.14}
\definecolor{richblack}{rgb}{0.0, 0.25, 0.25}
\definecolor{anti-flashwhite}{rgb}{0.95, 0.95, 0.96}
\usepackage{subcaption}
\usepackage{algorithm}
\usepackage{algpseudocode}
\usepackage{tikz}
\usepackage{listings}
\usepackage{multirow}
\usepackage{hhline}
\usepackage{makecell}
\usepackage{amsmath}
\usepackage{amsfonts}
\usepackage{url}

\usepackage{pgfplots}
\usepackage{svg}

\def\BibTeX{{\rm B\kern-.05em{\sc i\kern-.025em b}\kern-.08em
    T\kern-.1667em\lower.7ex\hbox{E}\kern-.125emX}}

\newcommand{\TerEffic}{{\textit{TerEffic}}}

\begin{document}

\title{TerEffic: Highly Efficient Ternary LLM Inference on FPGA}
\author{
    \IEEEauthorblockN{
        Chenyang Yin\IEEEauthorrefmark{1}, 
        Zhenyu Bai\IEEEauthorrefmark{2}, 
        Pranav Venkatram\IEEEauthorrefmark{2}, 
        Shivam Aggarval\IEEEauthorrefmark{2}, 
        Zhaoying Li\IEEEauthorrefmark{2}, 
        and Tulika Mitra\IEEEauthorrefmark{2}
    }
    \IEEEauthorblockA{
        \IEEEauthorrefmark{1}School of Electronic Engineering and Computer Science, Peking University\\
        Email: ycy@stu.pku.edu.cn
    }
    \IEEEauthorblockA{
        \IEEEauthorrefmark{2}School of Computing, National University of Singapore\\
        Email: zhenyu.bai@nus.edu.sg\\
        Email: e0552200@u.nus.edu\\
        Email: shivam@comp.nus.edu.sg\\
        Email: zhaoying@comp.nus.edu.sg\\
        Email: tulika@comp.nus.edu.sg
    }
}
\maketitle

\begin{abstract}
Deploying Large Language Models (LLMs) efficiently on edge devices is often constrained by limited memory capacity and high power consumption. Low-bit quantization methods, particularly ternary quantization, have demonstrated significant potential in preserving model accuracy while substantially decreasing memory footprint and computational costs. However, existing general-purpose architectures and accelerators have not fully exploited the advantages of low-bit quantization due to insufficient specialized hardware support.

We introduce TerEffic, an FPGA-based architecture tailored for ternary-quantized LLM inference. The proposed system offers flexibility through reconfigurable hardware to meet various system requirements. We evaluated two representative configurations: a fully on-chip design that stores all weights within on-chip memories, scaling out using multiple FPGAs, and an HBM-assisted design capable of accommodating larger models on a single FPGA board.

Experimental results demonstrate significant performance and energy efficiency improvements. For single-batch inference on a 370 M-parameter model, our fully on-chip architecture achieves 16,300 tokens/second, delivering a throughput 192× higher than NVIDIA Jetson Orin Nano with a power efficiency of 455 tokens/second/W, marking a 19× improvement. The HBM-assisted architecture processes 727 tokens/second for a larger 2.7B-parameter model—3× the throughput of NVIDIA A100—while consuming only 46W, resulting in a power efficiency of 16 tokens/second/W, an 8× improvement over the A100.


\end{abstract}


\section{Introduction}
\label{sec:intro}
Large Language Models (LLMs) have revolutionized AI by delivering unprecedented performance across diverse applications. State-of-the-art LLMs such as ChatGPT~\cite{chatgpt}, Claude~\cite{claude}, and Google Gemini~\cite{gemini} continue scaling upwards in parameter count to enhance their capabilities. However, a concurrent and significant trend has emerged toward more efficient inference through the deployment of smaller yet highly effective models. This efficiency is primarily achieved through advanced techniques like model distillation~\cite{kd, deepseek} and low-bit quantization~\cite{awq, quip, bitnet, bitnet1.58}. For example, Meta’s Llama3.2~\cite{llama3.2} provides lightweight models with 1B and 3B parameters specifically optimized for high-performance edge tasks, while GPT-4o mini~\cite{gpt4omini} demonstrates superior capabilities to GPT-4 at significantly reduced inference costs.

\begin{figure}[h]
    \centering
    \includegraphics[width=.9\linewidth]{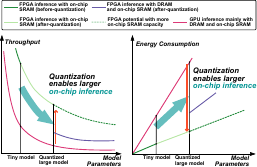}
    \caption{On-chip vs. Off-chip Memory for Inference: Throughput and Energy Trends with Increasing Model Parameters}
    \label{fig:insight}
    \vspace{-3mm}
\end{figure}

Nevertheless, existing AI inference accelerators—such as GPUs, NPUs, and TPUs—do not fully align with this emerging trend. Firstly, these accelerators lack specialized hardware tailored to efficiently handle the arithmetic operations and data movement patterns specific to low-bit quantized models, particularly binary and ternary LLMs. Secondly, current accelerators are predominantly DRAM-centric, relying heavily on off-chip High Bandwidth Memory (HBM). Such DRAM-based designs incur considerable energy overhead and suffer from bandwidth bottlenecks due to the memory wall~\cite{memory_wall}. However, aggressive model distillation combined with low-bit quantization (e.g., 1-bit LLMs) significantly reduces memory requirements, practically enabling fully SRAM-based on-chip inference. As illustrated in Figure~\ref{fig:insight}, low-bit quantization allows larger models to fit entirely within on-chip SRAM, making architectures with distributed SRAM—such as FPGAs—particularly advantageous. By leveraging high on-chip memory bandwidth, FPGAs can substantially improve throughput and energy efficiency compared to conventional GPU-based architectures.

Therefore, we propose \TerEffic, a specialized FPGA-based accelerator architecture optimized explicitly for efficient ternary LLM inference. {\TerEffic} provides tailored hardware units designed for efficient ternary arithmetic with optimized memory design and datapaths to unlock the potential of low-bit quantized LLM inference. In addition, {\TerEffic} features versatility for ternary LLMs with different sizes and activation functions. 
To make our design practical under different system requirements, we propose two architecture variants:
\textbf{(a)} a \textbf{fully on-chip design} that stores all weights within on-chip memories and can scale out using multiple FPGAs through the GTY/QSFP interface for FPGA-FPGA direct data transfer.
\textbf{(b)} an \textbf{HBM-assisted design} that stores weights off-chip, capable of accommodating larger models on a single FPGA board. The HBM-assisted variant can be deemed as a specialized design that leverages the potential of ternary quantization while the fully on-chip design provides further improvement with full-SRAM inference.

Both designs demonstrate remarkable advantages over GPUs in terms of throughput and power efficiency. For a 370M-parameter model, our fully on-chip architecture achieves a single-batch throughput of 16,300 tokens/second ($192\times$ higher) and a power efficiency of 455 tokens/second/W ($19\times$ better) compared to NVIDIA's Jetson Orin Nano. For larger models that far exceed the on-chip memory capacity of the FPGA, our HBM-assisted architecture also shows high performance. For a 2.7B-parameter model, it provides a single-batch throughput of 727 tokens/second and a power efficiency of 16 tokens/second/W, which is $3\times$ and $8\times$ better compared to NVIDIA's A100, respectively.
\section{Background \& Related work}
\subsection{Binary and Ternary Quantizations}
\label{background}
Quantization has been a key strategy to accelerate inference as neural networks grow rapidly in parameter count. Among the various quantization methods, binary and ternary quantizations feature extremely compressed weights that require little memory capacity and bandwidth. They were initially validated in CNN models such as BNN \cite{BNN}, XNOR-Net \cite{Xnor}, and TWN \cite{TWN}, achieving high efficiency with minimal accuracy loss.

Entering the era of LLMs, larger model sizes and limited hardware resources further necessitate low-bit quantization, especially in the context of edge devices. LLMs utilizing binary or ternary quantization are also referred to as '1-bit' LLMs, as each weight theoretically takes less than 2 bits for storage. As a pioneer of 1-bit LLM, BitNet\cite{bitnet} employed binary quantization and introduced the BitLinear layer as a replacement for linear projections. As its variant and improvement, BitNet 1.58b\cite{bitnet1.58} utilized ternary quantization to match the performance of full-precision (fp16 or bf16) Transformers. Later, TernaryLLM\cite{ternaryllm} leveraged knowledge distillation in its training and surpassed previous low-bit quantization methods. However, as the Q, K, V matrices in the self-attention mechanism remain in high precision, the high-precision matrix multiplications were not eliminated. To address this issue and realize a fully 'MatMul-free' model, \cite{scalable} used an RNN variant to replace the traditional self-attention in their ternary model and achieved excellent performance. Furthermore, this 1-bit quantization trend was extended to the Vision Transformer (ViT) domain \cite{vit1.58} and a scaling law \cite{1bittheory} was established, providing a theoretical basis for the impressive performance of these 1-bit models. Advances in post-training reparameterization \cite{shiftaddllm} also paved the way for low-cost training of 1-bit LLMs. More recently, Microsoft released an official CPU inference framework for BitNet, bitnet.cpp\cite{bitnetcpp}. However, this framework focused on the software level, while ternary models still have significant potential for improvements from custom hardware.

       \subsection{FPGA-based Transformer Accelerators}
Field-programmable gate arrays (FPGAs) are widely used for accelerating various tasks.
In the context of deep learning, several works have contributed toward accelerating Transformer inference using FPGAs. Many early attempts \cite{DFX,AFastandFlexible,NPE,FTRANs,EfficientMethodsforMapping,AcceleratingTransformer-based,AccommodatingTransformer} adopted overlay architectures that utilized an instruction set and universal computing engine for all layers. While overlay designs demonstrate flexibility for different models, they always suffer from frequent data movements of intermediate results.  Later, some spatial architectures have been proposed, featuring specialization of distinct PEs for specific operations. \cite{spatial} systematically analyzed the potential for FPGA-based spatial LLM acceleration and built an custom HLS kernel library. FlightLLM\cite{flightllm} utilized the FPGA-specific DSPs and hierarchical memory. It introducd a sparse DSP chain and an always-on-chip decode scheme to reduce memory overhead, providing 55 tokens/second for a 7B LLM on Xilinx Alveo U280. Most recently, EdgeLLM\cite{edgellm} proposed a CPU-FPGA heterogeneous acceleration system. Highlighting a unified data format and customized FP16*INT4 computational units, it outperformed the performance of \cite{flightllm} by 10\%-20\%. These works have demonstrated the potential for efficient FPGA LLM acceleration, yet they didn't realize fully on-chip inference due to the limited on-chip memory capacity, and therefore suffered from the memory bandwidth bottleneck.

While fully on-chip inference on FPGAs has been minimally explored, 1-bit quantization offers a promising avenue for this efficient paradigm. FPGAs provide an ideal platform to harness the advantages of 1-bit LLMs, thanks to their bit-level reconfigurability that enables customized memory architectures and computational units. While \cite{scalable} proposed a basic FPGA implementation for their ternary model, their prototype lacked effective hardware optimizations and failed to take full advantage of 1-bit quantization. 

\section{{\TerEffic} Architecture}
\begin{figure}[h]
    \centering
    \includegraphics[width=\linewidth]{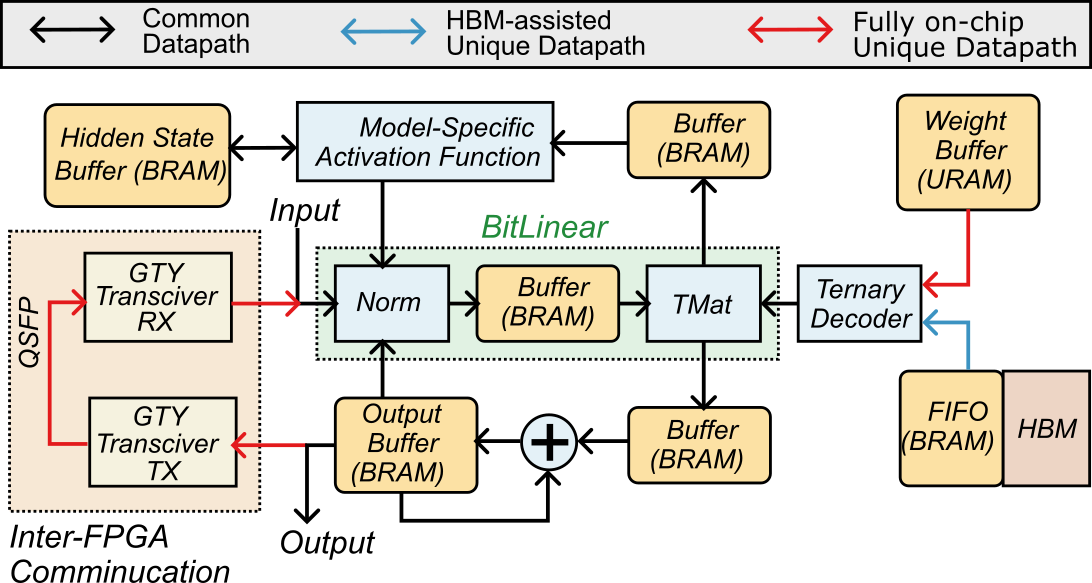}
    \caption{Architecture Overview}
    \label{fig:arch}
    \vspace{-3mm}
\end{figure}
\subsection{Architecture Overview}
\label{sec:archi}
The architecture of {\TerEffic} is shown in Figure \ref{fig:arch}.
One key design to fit a ternary model with minimal memory is the compression of ternary weights. We use 1.6-bit compression in weight memory and restore them to 2 bits in the Ternary Decoder. The specific details will be explained in Section \ref{1.6bit}. The BitLinear Module is the most important module in the architecture, consisting of the RMSNorm Module, the TMat (Ternary Matrix Multiplication) Core, and the activation buffer between them.  The module details will be provided in Section \ref{RMSNorm} and \ref{TMat}. In addition, a reconfigurable module executes the activation function, connecting to a hidden state buffer that stores the hidden state vector from the previous timestep. 

To improve practical usability, we propose two architecture variants with different memory hierarchies: the \textbf{fully on-chip} variant stores all weights on-chip and scales out with multiple FPGAs and the FPGA-FPGA direct data transfer, while the \textbf{HBM-assisted} variant uses off-chip HBM for incorporating larger models on a single FPGA. The black lines in Figure \ref{fig:arch} represent the common datapath shared by the two variants, while the red and blue lines represent the datapath unique to the fully on-chip architecture and the HBM-assisted architecture, respectively. The details of the two architecture variants will be discussed in Section \ref{sec: memory}.

\subsection{1.6-Bit Weight Compression}
\label{1.6bit}
\begin{figure}
    \centering
    \includegraphics[width=.8\linewidth]{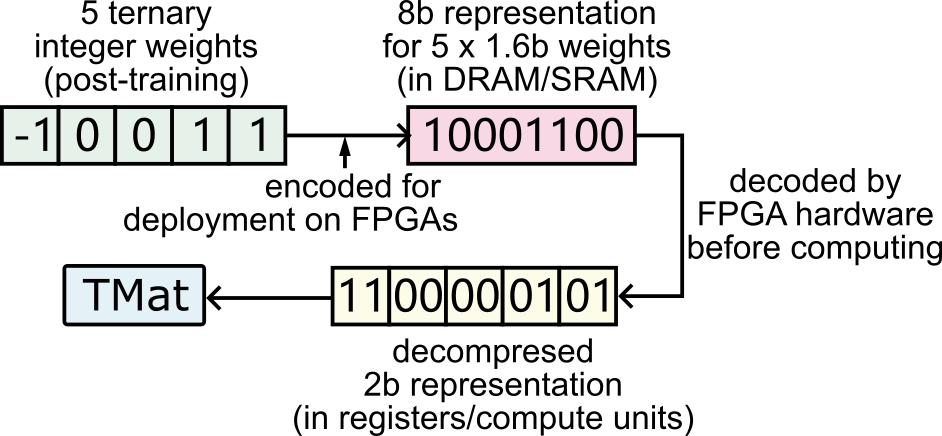}
    \caption{1.6-Bit Weight Compression}
    \label{fig:bit-compression}
    \vspace{-7mm}
\end{figure}
Ternary weights \{-1, 0,1\} have a theoretical storage of 1.58 bits/weight that is not natively supported by digital circuits. A straightforward 2-bit representation causes $0.42/2 \approx 20\%$ of storage waste. To overcome this challenge, we propose a 1.6-bit encoding/decoding scheme that reduces the storage waste, inspired by~\cite {ternaryencoding}. Figure~\ref {fig:bit-compression} details the compression scheme: we observed that 5 ternary weights have 243 possible combinations, whereas 8 binary bits can represent 256 distinct values. Hence, 5 ternary weights can be encoded using 8 bits, averaging 1.6 bits per weight. For example, five original weights \{-1,0,0,1,1\} are encoded into 8 bits (10001100) when stored in memory. This encoding is performed after the quantization of the model. During the inference, the 8b-encoded weights are decoded into $5\times2b$ representations by the Ternary Decoder. The stored weights (in 1.6-bit format) are thereby decoded into 2-bit format where 01 corresponds to 1, 11 to -1, and 00 to 0. As the decoding involves only bitwise operations, such as + and \&, it incurs minimal hardware cost and latency.

\subsection{RMSNorm Module}
\label{RMSNorm}
\begin{figure}[h]
    \centering
        \includegraphics[width=0.8\linewidth]{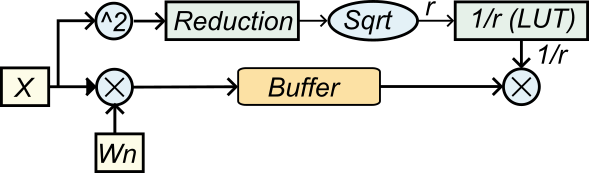} 
        \caption{RMSNorm Module}
        \label{fig:RMSNorm}
        \vspace{-5mm}
\end{figure}

The RMSNorm\cite{RMSNorm} is more computationally efficient than the traditional LayerNorm\cite{attention} while maintaining high accuracy, making it well-suited for FPGA. The algorithm for RMSNorm is presented below:
\begin{equation}
        r = \sqrt{\frac{1}{d} \sum_{i=1}^d x_i^2 + \epsilon} \quad , \quad 
        \text{RMSNorm}(X) = \frac{X\odot W_n}{r}
\label{eq:RMSNorm}
\end{equation}

where $X\in\mathbb{R}^{1 \times d}$ denotes the input activation, $W_n\in\mathbb{R}^{1 \times d}$ denotes the normalization weight, $r$ denotes the Root-Mean-Square(RMS) result , $\epsilon$ is a small constant and $\odot $ represents dot product. As shown in the architecture in Figure \ref{fig:RMSNorm}, the RMS computation and $X\odot W_n$ can be executed in parallel. As the RMS process has longer latency, the $X\odot W_n$ results are temporarily stored in SRAM-based on-chip buffers. Moreover, as divisions incur high hardware cost and long cycle latency, we replace the expensive divisions ($\div r$) with DSP-based multiplications ($\times \frac{1}{r}$), using r as an index to retrieve 1/r from an on-chip look-up table consisting of a small amount of SRAMs. This hardware optimization saves hardware resources and allows for a higher frequency.

\subsection{Ternary Matrix Multiplication (TMat) Core}
\label{TMat}
Matrix multiplications constitute the dominant computational workload (65\%-85\%) in typical LLMs \cite{high_perf}. To fully harness the potential of ternary LLMs, we introduce a specialized hardware unit for ternary matrix multiplications, called the \textit{TMat Core}.

The TMat Core serves as the primary computational engine for ternary matrix multiplications ($X\times W = X'$), where $X \in \mathbb{R}^{1 \times d}$ is the normalized input activation vector (int8), $W \in \mathbb{R}^{d \times d'}$ is the ternary weight matrix after decoding (2 bit/weight when entering compute units, as illustrated in Figure~\ref{fig:bit-compression}), and $X' \in \mathbb{R}^{1 \times d'}$ represents the output activation (int8). The TMat Core has a fixed dimension ($d = d' = 256$, i.e., $X \in \mathbb{R}^{1 \times 256}$ and $W \in \mathbb{R}^{256 \times 256}$), and tiling is applied to accommodate different matrix multiplication sizes.

\begin{figure}
        \centering
        \includegraphics[width=\linewidth]{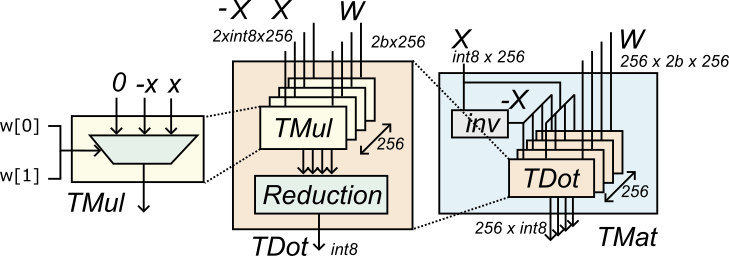}  
        \caption{Ternary Matrix Multiplication(TMat) Core}
        \label{fig:TMM}
        \vspace{-5mm}
\end{figure}

As depicted in Figure~\ref{fig:TMM} and Listing~\ref{lst:tmat}, the TMat Core employs a hierarchical structure consisting of 256 TDot units. Each TDot unit calculates a 256-point dot product between the activation vector ($X$) and a ternary weight vector, which is a column in the weight matrix ($W$). Further decomposing the TDot unit, it contains 256 TMul units, each performing a multiplication between an 8-bit integer activation element($x$) and a 2-bit ternary weight($w$), and a reduction unit that sums these 256 products into a partial sum every cycle. The partial sums are accumulated in an accumulator, ultimately forming one output activation element. The 256 weight columns are stored contiguously in one row of weight memory (either on-chip URAM or HBM depending on the specific design variant, as shown in figure~\ref{fig:arch}). These weights are fetched, decoded, and distributed to the respective TDot units. The whole TMat Core is composed of LUTs, the basic logic units in FPGAs. Compared to previous FPGA-based accelerators like \cite{flightllm} and \cite{edgellm} that utilize power-costly DSPs for matrix multiplications, the LUT-based TMat Core leverages the advantages of ternary quantization.

Given that the TMat Core computes vector-matrix multiplications, the input activation vector is reused across all TDot units. Since a ternary multiplication($x*w, x \in X,w \in W$) outcome is limited to $-x$, $x$, or $0$, we optimize the architecture by precomputing the inversions of the activation elements($-x$) within the TMat unit, distributing both the activation and its inversion($X$ and $-X$) to all TDot units. 
Experiments show a 13.2\% reduction in resource utilization (LUT usage) by applying this optimization, compared to a basic implementation with individual inversion logic in each TMul unit.

\begin{lstlisting}[float, caption=Optimized TMat Core Computation, label={lst:tmat},basicstyle=\scriptsize\ttfamily]
// W[256][256] is stored in one row in the weight memory, fetched and decoded simultaneously before going into the TMat Core
fn TMat(W: 2b[256][256], X: int8[256]){
    O : int8[256][256]; // TMul outputs
    S : int8[256]; // TDot partial sums
    nX : int8[256] // inversion of X
    parallel_for i in 0..255{
        nX[i]=-X[i]; // precompute the inversions
    }  
    parallel_for i in 0..255 { // For each TDot
        parallel_for j in 0..255 { // For each TMul
            if W[i][j] == '01'
                O[i][j] = X[i];
            else if W[i][j] == '11'
                O[i][j] = nX[i];
            else if W[i][j] == '00'
                O[i][j] = 0;
        }
        // performed by the reduction unit
        S[i] = sum(j in 0..255, O[i][j]);
    }
}
\end{lstlisting}

Here is an example of the tiling for larger matrix multiplication sizes. Consider a 1024×1024 ternary matrix multiplication, where $X\in\mathbb{R}^{1 \times 1024}$ and $W\in\mathbb{R}^{1024 \times 1024}$. With a tiling factor of 256 matching the TMat Core, it takes 4 cycles to process a complete activation input, completing the dot products between the activation and 256 columns of weights and generating 256 output elements. To produce the full output $X'\in\mathbb{R}^{1 \times 1024}$, the activation input is repeated four times. Including an additional 8 cycles for the reduction step. The total latency of the TMat Core becomes $4\times4 + 8 = 24$ cycles.

\section{{\TerEffic} Memory Architecture}
\label{sec: memory}

\subsection{Compute-Memory Alignment}
\label{sec:Compute-Memory Alignment}

FPGA platforms typically offer multiple on-chip memory types, each characterized by distinct bandwidth and capacity trade-offs. The two on-chip SRAM types on the AMD U280 FPGA~\cite{amd_u280} are shown in Table~\ref{tab:SRAM}, where Cap and Bw denote memory capacity and bandwidth, respectively. A higher bandwidth per capacity means this memory is more suitable for tasks with high bandwidth requirements. Specifically, BlockRAMs (BRAMs) deliver higher bandwidth per capacity but have limited total storage capacity, whereas UltraRAMs (URAMs) provide larger capacity at the expense of lower bandwidth per capacity. 

\begin{figure}[h]
    \centering
    \includegraphics[width=.8\linewidth]{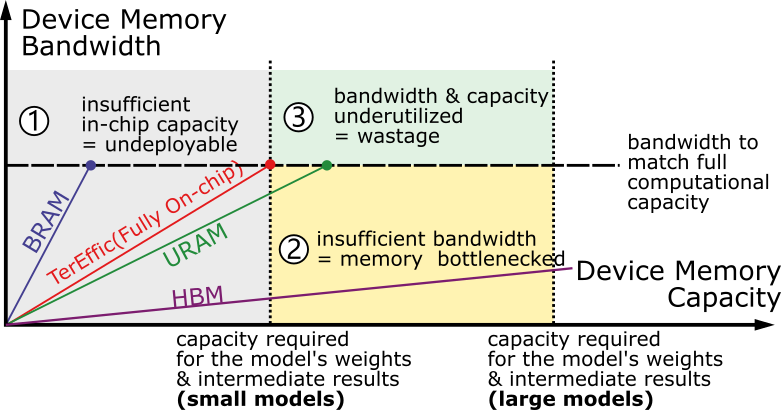}
    \caption{Compute-Memory Alignment}
    \label{fig:Alignment}
    \vspace{-4mm}
\end{figure}

To achieve maximum utilization of computational resources, the memory hierarchy must align closely with both bandwidth and capacity requirements dictated by the model. Figure~\ref{fig:Alignment} illustrates the high-level design space for memory system configurations, highlighting three distinct zones:
Zone~\textcircled{1} represents configurations that lack sufficient capacity, typical for purely BRAM-based designs. Zone~\textcircled{2} represents configurations limited by insufficient bandwidth, commonly seen in fully URAM-based solutions. Zone~\textcircled{3} indicates scenarios where both memory bandwidth and capacity exceed the requirements, resulting in memory resource underutilization. An optimal design lies at the intersection of these three zones, effectively balancing bandwidth and capacity to fully utilize available computational resources. To achieve this balance, we propose a hybrid fully on-chip architecture (orange line in Figure~\ref{fig:Alignment}). In this design, weights are primarily stored in URAMs due to their higher capacity, while activations and intermediate computations leverage the high-bandwidth BRAMs as buffers. This approach allows our architecture to closely align with both bandwidth and capacity requirements, optimizing resource utilization and energy efficiency.

\begin{table}[h]
    \centering
    \caption{Attibutes of On-chip Memory Types on U280}
    \label{tab:SRAM} 
    \resizebox{\linewidth}{!}{
    \begin{tabular}{|r|r|r|r|r|r|}
        \hline
        Catagory& Cap/piece & Bw/piece & Bw/Cap & Number & Total Capacity \\
        \hline
        BRAM & 36Kb & 72b & 0.002 & 2,016 & 8.85MB \\
        \hline
        URAM & 288Kb & 144b & 0.0005 & 960 & 33.75MB \\
        \hline
    \end{tabular}
    }
    \vspace{-5mm}
\end{table}

With low-bit quantization such as ternary quantization, small models enable all weights to be fully stored in on-chip memory, achieving high performance by leveraging the high on-chip SRAM bandwidth. However, as the model size grows, the available on-chip memory capacity inevitably becomes insufficient. In such cases, utilizing an off-chip HBM with large capacity becomes necessary. Nonetheless, due to its comparatively lower bandwidth relative to on-chip SRAM, using HBM may introduce performance bottlenecks, as illustrated by the purple line in Figure~\ref{fig:Alignment}.

To address these varying constraints, we propose two architecture variants: the \textbf{fully on-chip} and the \textbf{HBM-assisted} architectures. The fully on-chip architecture prioritizes short latency but is constrained to smaller models, requiring either FPGAs with larger on-chip memory or scaling across multiple cards to provide sufficient memory capacity for larger models. Conversely, the HBM-assisted architecture prioritizes throughput, leveraging batch-parallelism to improve weight reuse, thereby aligning the HBM bandwidth with computational capabilities.

\subsection{Fully On-chip Architecture}
\label{on-chip}
The fully on-chip architecture is designed to store all model weights within the FPGA’s on-chip memory. For instance, the AMD U280 FPGA provides approximately 42MB of on-chip memory, capable of storing around 210M ternary weights(1.6 bits/weight). In a single-card deployment, this fully on-chip architecture already offers the best possible performance by exclusively leveraging tensor-parallelism, i.e. exploiting parallelism within the TMat Core. 
However, in practice, the smallest LLM that demonstrates acceptable accuracy on standard benchmarks has approximately 400M parameters—for example, the 370M parameter model presented in~\cite{scalable}—which exceeds the on-chip memory capacity of most current FPGAs. To address this limitation and enhance the practicality of our architecture, we propose a multi-FPGA deployment technique where different layers of the model are distributed across the on-chip memory of multiple cards (layer-parallelism).

\begin{figure}[h]
    \centering
    \includegraphics[width=.9\linewidth]{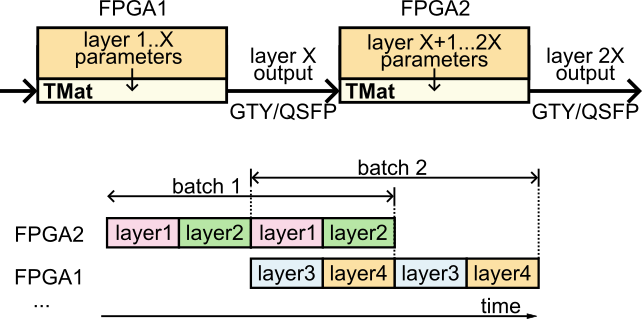}
    \caption{Layer-Parallelism for Fully On-chip Inference with Multiple Cards}
    \label{fig:fully-on-chip-mult-cards}
    \vspace{-5mm}
\end{figure}

As depicted in Figure~\ref{fig:fully-on-chip-mult-cards}, this multi-card system employs pipeline-parallelism across layers. In an M-card deployment, M batches are processed concurrently, with each FPGA executing a different batch at distinct pipeline stages simultaneously. Consequently, this pipeline-parallelism results in an approximately M-fold increase in throughput compared to a naive sequential execution approach.

The FPGA-FPGA communication utilizes the GTY transceiver interface, connected to QSFP28 sockets provided on the U280 board. Each FPGA uses eight GTY channels for two 4-lane QSFP28 sockets, acquiring up to 200 Gbps bandwidth~\cite{amd_u280}. The FPGAs are directly interconnected using cables, arranged in accordance with the sequential order of layers.


\subsection{HBM-assisted Architecture}
\label{HBM}
For even larger models that require an impractical number of FPGAs, we propose the HBM-assisted architecture. The HBM on U280 is able to store up to 8GB of data off-chip, approximately the size of a 40B ternary-quantized model. As any architecture that relies on the off-chip memory, such an architecture can be easily memory-bottlenecked: For a single-batch task, the HBM is not able to provide enough bandwidth to keep the compute core (the TMat Core) busy. 

\begin{figure}[h]
    \centering
    \includegraphics[width=.95\linewidth]{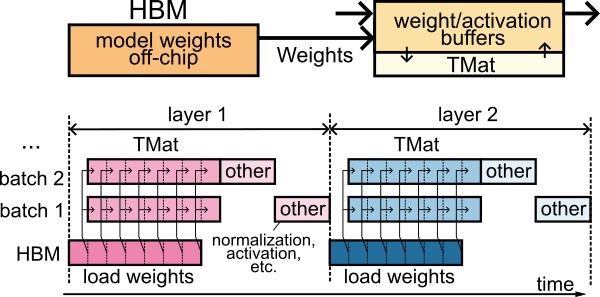}
    \caption{Batch-Parallelism for HBM-assisted Architecture}
    \label{fig:hbm-version}
    \vspace{-5mm}
\end{figure}

To alleviate the bottleneck, batch-parallelism is leveraged to improve the reuse of the weights loaded from HBM. As shown in Figure~\ref{fig:hbm-version}, multiple batches are processed simultaneously in the TMat Core, with the ternary weights shared. For an N-batch task, the TDots in the TMat core are divided into N groups, each group executing one batch. 

\begin{figure}[h]
    \hspace{8mm}
    \includegraphics[width=.7\linewidth]{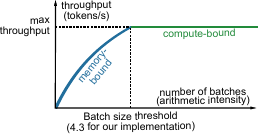}
    \caption{Roofline Model for HBM-assisted Architecture}
    \label{fig:roofline}
    \vspace{-3mm}
\end{figure}


To evaluate the performance under different batch sizes, we refer to the Roofline model\cite{roofline} depicted in Figure \ref{fig:roofline}. 
As the batch size increases, the total number of computation operations grows proportionally, whereas the volume of weights transferred from HBM remains constant as the same weights are shared across all batches. This leads to a rise in arithmetic intensity, defined as the ratio of computation operations to memory access. According to the Roofline model, this shift in arithmetic intensity causes the transition from a memory-bound regime to a compute-bound regime.

As illustrated in Figure~\ref{fig:roofline}, when operating in the memory-bound region, the system throughput increases with the batch size due to the effective reuse of weights. Once the arithmetic intensity and the batch size exceed a threshold, the architecture enters the compute-bound region, where the throughput saturates at its maximum achievable level and remains constant despite further increases in batch size. In our actual implementation, the batch size threshold is 4.3. Therefore, for tasks with batch sizes bigger than 5, the system becomes compute-bound and the throughput reaches the maximum. 





To realize batch-parallelism, we need N activation buffers between the RMSNorm Module and the TMat Core to input N activations simultaneously. In our actual implementation, we set 16 buffers for the 16-batch task. Due to the routing congestion issue, we keep the 16 buffers for the single-batch task to reduce the fan-in and fan-out count. Therefore, the architecture remains unchanged for single-batch and multi-batch evaluations. In addition, the HBM loading is asynchronous from the compute core, with HBM running at 450 MHz and the core at 150 MHz. A BRAM-composed FIFO buffer is thereby inserted between the HBM and the core computing logic. Concretely, on the U280 board, all 32 HBM channels are used at 450 MHz, providing a peak bandwidth of 460 GB/s.

\section{Evaluation}
\subsection{Evaluation Methodology}
\subsubsection{Demonstration Ternary Models}
\begin{figure}[h]
    \centering
    \includegraphics[width=.9\linewidth]{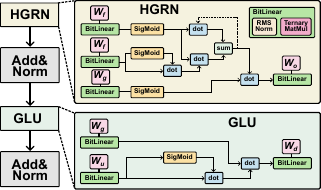}
    \caption{MatMul-free LM \cite{scalable} Layer Computation Graph}
    \label{fig:model}
    \vspace{-4mm}
\end{figure}

We use the MatMul-free LM models~\cite{scalable} as the demonstration ternary models. 
Figure~\ref{fig:model} shows the layer computation graph. The models have been proven to match the performance of the similar-size SOTA Transformers, while eliminating high-precision matrix multiplications~\cite{scalable}. Analogy to traditional Transformers~\cite{attention}, the HGRN~\cite{HGRN} is an RNN-based alternative to the self-attention mechanism, while the GLU~\cite{glu}, widely used in models like Llama\cite{llama}, is recognized by many as a robust enhancement for feed-forward networks (FFNs). The model-specific activation functions (introduced in Section \ref{sec:archi}) include addition, subtraction and dot product (Dot) that can be directly implemented using LUTs and DSPs, while the sigmoid function is implemented by a BRAM-based look-up table.

Table~\ref{tab:models} shows different attributes of the models for evaluation. The model parameters range from 370M to 2.7B: we evaluate the fully on-chip architecture using the 370M model and the HBM-assisted architecture using the larger models.
\begin{table}[h]
    \centering
    \caption{Attributes of the Demonstration Models}
    \label{tab:models} 
    \begin{tabular}{|r|r|r|r|}
        \hline
        Parameter& Dimension & Layer & Storage Size  \\
        \hline
        370M & 1024 & 24 & 58MB \\
        \hline
        1.3B & 2048 & 24 & 230MB \\
        \hline
        2.7B & 2560
        & 32 & 480MB \\
        \hline
    \end{tabular}
    \vspace{-3mm}
\end{table}
\subsubsection{Baselines}
To demonstrate the effectiveness of our architectures, we first compare against the basic FPGA implementation from~\cite{scalable} as our FPGA baseline. Additionally, to emphasize the performance advantages of our design, we include GPU baselines by deploying the 370M model on an NVIDIA edge GPU, Jetson Orin Nano, optimized using TensorRT~\cite{TensorRT} for maximum inference performance. Since larger models (with 1.3B and 2.7B parameters) exceed the memory capacity of the Jetson Orin Nano, these models are deployed on NVIDIA A100 GPU, again utilizing TensorRT optimization. GPU baselines are quantized to the lowest supported precision: INT8 for the Jetson Orin Nano and INT4 for the A100. The hardware specification of the baselines and ours is listed in Table \ref{tab:specs}.

It is worth noting that {\TerEffic} specifically accelerates the decoder layers, whereas our GPU baseline experiments include the other layers in the entire model (embedding, position encoding, and output layers). However, additional GPU experiments indicate that these extra layers account for less than 0.1\% of the total computational workload and inference runtime on GPUs. Thus, the throughput comparisons presented remain fair and representative of actual performance differences.

\begin{table}[h]
    \centering
    \caption{Hardware Specification of Baselines and {\TerEffic}}
    \resizebox{\linewidth}{!}{
    \begin{tabular}{|c|c|c|c|}
        \hline
                   & Tech Node & Off-chip bw. & On-chip SRAM \\
                   \hline
         \makecell{AMD Alveo U280\\(Ours \TerEffic)} & 16 nm & 460GB/s & 42 MB \\
         \hline
         \makecell{Intel PAC D5005\\(\cite{scalable})} & 14 nm & 77GB/s & 30.5 MB \\
         \hline
         NVIDIA Jetson Orin & 8 nm & 68.29GB/s & (L1+L2) 1.25 MB \\
         \hline
         NVIDIA A100 & 7 nm & 1935GB/s & (L1+L2) 100.25 MB \\
         \hline
    \end{tabular}
    }
    \vspace{-5mm}
    \label{tab:specs}
\end{table}

\subsection{{\TerEffic} Implementation Details}
\label{sec:fpga-impl}
\begin{figure}[h]
    \centering
    \begin{subfigure}[b]{0.2\textwidth}
        \centering
        \includegraphics[width=\textwidth]{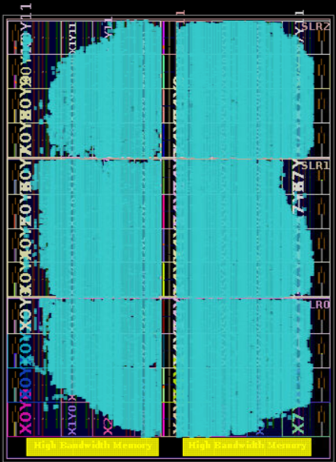}
        \caption{Fully On-Chip Arch}
    \end{subfigure}
    \hspace{5mm}
    \begin{subfigure}[b]{0.2\textwidth}
        \centering
        \includegraphics[width=\textwidth]{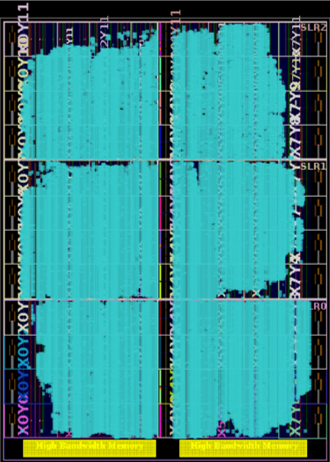}
        \caption{HBM-Assisted Arch}
    \end{subfigure}
    \caption{Layouts of the Two Architectures}
    \label{fig:layout}
    \vspace{-4mm}
\end{figure}

We performed synthesis, placement, and routing on Vivado v2023.2, achieving \textbf{150MHz} frequency. The layouts of the two architectures are presented in Figure \ref{fig:layout}. 

\begin{table}[h]
    \centering
    \caption{Resource Utilization (Unit Numbers and Percentage) for the Fully On-chip  and the HBM-assisted Architecture}
    \resizebox{\linewidth}{!}{
    \begin{tabular}{|c|c|c|c|c|}
         \hline
         Resource& Fully On-chip Arch& (\%) & HBM-assisted Arch & (\%) \\
        \hline
        LUT& 781K &59.88 & 760K &58.31\\
        \hline
        Register& 537K &20.61& 538K & 20.64\\
        \hline
        Carry8 & 53K & 32.51 & 53K & 32.52\\
        \hline
        DSP & 3,041 & 33.70 & 3,041 & 33.70\\
        \hline
        BRAM & 964& 47.82& 1669.5 & 82.81\\
        \hline
        URAM & 740 & 77.08 & 0 & 0 \\
        \hline
        QSFP & 8 & 100 & 0 & 0\\
        \hline
        HBM & 0 &0 &32 &100\\
        \hline
    \end{tabular}
    }
    \vspace{-5mm}
    \label{tab:resource-fully-on-chip}
\end{table}

\subsubsection{Resource Utilization} Table~\ref{tab:resource-fully-on-chip} shows the resource utilization in the two architectures, where 'QSFP' and 'HBM' refer to the channels.
As our FPGA design computes the ternary matrix multiplications with the customized TMat Core entirely composed of LUTs, the LUT utilization is significantly higher than that of costly DSPs. In the fully on-chip architecture, URAMs are utilized for weight storage, while BRAMs are used for activation buffers and look-up tables for functions like sigmoid and division. Therefore, URAM utilization is higher than BRAM utilization. All 8 QSFP28 channels are utilized for inter-card activation transfer, providing the maximum bandwidth of 200 Gbps. In the HBM-assisted architecture, all 32 HBM channels are fully utilized for the maximum HBM bandwidth of 460 GB/s. The BRAM utilization is higher than that in the fully on-chip architecture, mainly because of the dual-clock FIFO between the HBM and the computational units for asynchronous memory operations under different frequencies (write under 450MHz and read under 150MHz).



\subsubsection{Power Breakdown}
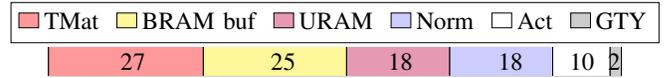
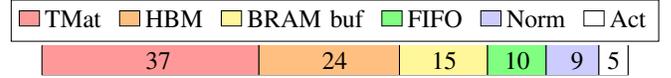
\begin{figure}
    \centering
    \begin{subfigure}{\linewidth} 
        \centering
    \resizebox{\linewidth}{!}{
\begin{tikzpicture}[]
    \begin{axis}[
    xbar stacked,
    width=1.1\linewidth,
    xmajorgrids = true,
    xmin=0,xmax=100,
    ytick style={draw=none},
    ytick = data, yticklabels = {},
    bar width=6mm, y=10mm,
    nodes near coords={}, 
    nodes near coords align={}, 
    enlarge y limits=0.5, 
    legend style={/tikz/every even column/.append style={column sep=0.18cm}},
    legend columns =6,
    legend style={at={(0.5,1.15)},anchor=south},
    xtick=\empty
    ]
        \addplot [fill=red!40,nodes near coords align={xshift=-10mm},nodes near coords={27}] coordinates{(27,1)};
        \addplot [fill=yellow!50,nodes near coords align={xshift=-9mm},nodes near coords={25}] coordinates{(25,1)};
        \addplot [fill=purple!40,nodes near coords align={xshift=-7mm},nodes near coords={18}] coordinates{(18,1)};
        \addplot [fill=blue!20,nodes near coords align={xshift=-6mm},nodes near coords={18}] coordinates{(18,1)};
        \addplot [fill=white!20,nodes near coords align={xshift=-4mm},nodes near coords={10}] coordinates{(10,1)};
        \addplot [fill=black!20,nodes near coords align={xshift=-1mm},nodes near coords={2}] coordinates{(2,1)};
        \legend{TMat,BRAM buf,URAM,Norm,Act, GTY}
    \end{axis}
\end{tikzpicture}
}
\caption{Dynamic Power Breakdown (\%) : Fully On-chip Architecture }
    \label{fig:on-chip-powerbreakdown}
    \end{subfigure}

    \vspace{4mm}
    
    \begin{subfigure}[b]{\linewidth} 
        \centering
    \resizebox{\linewidth}{!}{
\begin{tikzpicture}[]
    \begin{axis}[
    xbar stacked,
    width=1.1\linewidth,
    xmajorgrids = true,
    xmin=0,xmax=100,
    ytick style={draw=none},
    ytick = data, yticklabels = {},
    bar width=6mm, y=10mm,
    nodes near coords={}, 
    nodes near coords align={}, 
    enlarge y limits=0.5, 
    legend style={/tikz/every even column/.append style={column sep=0.18cm}},
    legend columns =6,
    legend style={at={(0.5,1.15)},anchor=south},
    xtick=\empty
    ]
        \addplot [fill=red!40,nodes near coords align={xshift=-14mm},nodes near coords={37}] coordinates{(37,1)};
        \addplot [fill=orange!50,nodes near coords align={xshift=-9mm},nodes near coords={24}] coordinates{(24,1)};
        \addplot [fill=yellow!50,nodes near coords align={xshift=-6mm},nodes near coords={15}] coordinates{(15,1)};
        \addplot [fill=green!50,nodes near coords align={xshift=-4mm},nodes near coords={10}] coordinates{(10,1)};
        \addplot [fill=blue!20,nodes near coords align={xshift=-3mm},nodes near coords={9}] coordinates{(9,1)};
        \addplot [fill=white!20,nodes near coords align={xshift=-2mm},nodes near coords={5}] coordinates{(5,1)};
        \legend{TMat,HBM, BRAM buf, FIFO,Norm,Act}
    \end{axis}
\end{tikzpicture}
}
\caption{Dynamic Power Breakdown (\%) : HBM-assisted Architecture }
    \vspace{-5mm}
    \label{fig:HBM-powerbreakdown}
    \end{subfigure}
    \label{fig:powerbreakdown}
    \caption{Dynamic Power Breakdown for the Two Architectures: BRAM buf denotes the BRAM buffers for activations, Act denotes the non-linear activation functions, and FIFO denotes the BRAM FIFO between HBM and the TMat Core.}
    \vspace{-5mm}
\end{figure}

From the Vivado power report, we obtained the single-card power (denoted as $P_0$) of the fully on-chip architecture: $P_0$ is \textbf{31.8W}, composed of static power ($P_{s}$) of 4.0W and dynamic power ($P_{d}$) of 27.8W. 
A detailed dynamic power breakdown is provided in Figure~\ref {fig:on-chip-powerbreakdown}, showcasing that the TMat Core is the most power-consuming unit in the architecture. Moreover, memory (BRAMs and URAMs) takes up 43\% power consumption, highlighting the significance of the hierarchical on-chip memory architecture.

The power of the HBM-assisted architecture is \textbf{46.2W}, with a detailed breakdown presented in Figure \ref{fig:HBM-powerbreakdown}. The introduction of HBM and the corresponding BRAM-based FIFOs leads to an increase in memory-related power consumption. The power consumption of the TMat Core also increases with a more severe routing congestion, mainly because of the high BRAM utilization due to the FIFOs.

\subsection{Fully On-chip Architecture Evaluation}
We evaluated the performance of our fully on-chip architecture using the 370M-parameter model. As detailed in Section \ref{on-chip}, each AMD U280 FPGA can hold approximately 210M ternary weights entirely within on-chip memory. Consequently, a two-card system is required, with each FPGA responsible for half of the model’s 24 layers.

Activations $(1024 \times 8b)$ are transferred between cards via eight QSFP28 channels, offering a combined bandwidth of 200 Gbps. This interconnect latency accounts for less than 0.1\% of the total inference latency. Operating at 150 MHz, the system achieves a single-batch throughput of approximately 16,300 tokens/second. For multi-batch tasks, both cards can operate in parallel using the layer-parallelism strategy described in Section \ref{on-chip}, doubling the throughput to 32,600 tokens/second. In the single-batch mode, only one FPGA is active, while the other remains idle and consumes only static power ($P_s$). The total system power in this case is $P_0+P_{s}=35.8W$. Under multi-batch execution, both cards are fully utilized, resulting in a total power consumption of $63.6 W$.



\begin{table}[h]
    \centering
    \caption{Comparison of {\TerEffic}(Fully On-Chip) with GPU and a Basic FPGA Design\cite{scalable} for a 370M Model.}
    \label{tab:onchip_comparison} 
    \resizebox{\linewidth}{!}{
    \begin{tabular}{|r|r|r|r|r|r|r|r|}
        \hline
        Hardware& Batch &TP & TP& Power& Eff &Eff \\
        & Size &(tk/s)& Impv& (W) & (tk/s/W) & Impv\\
        \hline
        \cite{scalable}(FPGA)  &  1&  62& 0.7$\times$ & 13.7 &5 & 0.2$\times$ \\
        \hline
        Jetson  &  1& 85& 1$\times$ & 3.5 &24 & 1$\times$ \\
        
        \hline
        \textbf{{\TerEffic} (Ours)} &  1&  16,300& 192$\times$&  35.8&455& $19\times$ \\
        \hline
        \hline
        Jetson  & 16&  1,076&1$\times$&  4.4 & 245& 1$\times$ \\
        \hline
        \textbf{{\TerEffic} (Ours)}  &  16&  32,600&$30\times$&
        63.6&513& $2.1\times$ \\
        \hline
    \end{tabular}
    }
\end{table} 
We compare our fully on-chip results of the 370M model with the FPGA results from \cite{scalable} and the experimental results on Jetson Orin Nano, with the GPU batch size set to 1 and 16 for single-batch and multi-batch tasks respectively. The comparison is presented in Table \ref{tab:onchip_comparison}, where 'TP' denotes throughput (tokens/second), 'Eff' denotes power efficiency (tokens/second/watt), and 'Impv' denotes the throughput or power efficiency improvement over Jetson Orin Nano. While the baseline ternary FPGA design is not competitive, {\TerEffic} significantly outperforms the edge GPU in terms of both throughput and power efficiency, achieving a $\mathbf{192\times}$ throughput and a $\mathbf{19\times}$ power efficiency under single-batch tasks. Under the 16-batch task, despite not being specially designed for big batch sizes, our design still exceeds the GPU's throughput and power efficiency by $\mathbf{30\times}$ and $\mathbf{2.1\times}$, respectively. 

According to a report from TSMC\cite{tsmc}, TSMC's 7nm technology delivers 65\% power reduction at the same speed compared to its 16nm technology node. Therefore, we roughly estimate our power efficiency to be $2\times$ better under the 8nm technology node(same as Jetson Orin Nano). Consequently, our projected single-batch and multi-batch power efficiency will be respectively $\mathbf{38\times}$ and $\mathbf{4.2\times}$ under the same technology node. The architecture can also serve as a principle for an ASIC ternary accelerator design, which will offer even better performance. These improvements highlight the exceptional efficiency of our fully on-chip architecture, as it fully utilizes the massive on-chip SRAM bandwidth compared to the GPU's limited off-chip DRAM bandwidth. Moreover, while GPUs lack low-precision support, our customized TMat Core unlocks substantial computational capabilities through efficient ternary operations.
\subsection{HBM-assisted Architecture Evaluation}
The HBM-assisted architecture is used to accelerate larger models, like the 1.3B and 2.7B models, with only a single card required. Following the Roofline model discussed in Section \ref{HBM}, the HBM-assisted architecture is memory-bound when the batch size and the arithmetic intensity are small. The single-batch throughput is thereby constrained by the HBM bandwidth, which is 1489 tokens/second for the 1.3B model and 727 tokens/second for the 2.7B model. When the batch size increases, we utilize the batch-parallelism, and the system will transform from memory-bound to compute-bound. Referencing Figure \ref{fig:roofline}, the multi-batch throughput will first increase with the batch size, and then saturate after the system becomes compute-bound. In our implementation, the batch size threshold is 4.3. Therefore, a 16-batch task delivers the maximum throughput of 5,885 tokens/second and 3,028 tokens/second for the 1.3B and 2.7B models, respectively.

We compare our results with the FPGA baseline\cite{scalable} on the 1.3B model, and conduct A100 experiments on the 1.3B and 2.7B models, with the batch size also set to 1 and 16. The results are presented in Table \ref{tab:HBM_comparison}. Similarly, our projected single-batch and multi-batch power efficiency is $16\times$ and $5\times$ under the same technology node of A100.

\begin{table}[h]
    \centering
    \caption{Comparison of {\TerEffic}(HBM-assisted) with GPU and a Basic FPGA Design\cite{scalable} for Larger Models}
    \label{tab:HBM_comparison} 
    \resizebox{\linewidth}{!}{
    \begin{tabular}{|r|r|r|r|r|r|r|r|}
        \hline
        Model, &Hardware &TP & TP& Power&Eff &Eff \\
        Batch Size& &(tk/s)& Impv& (W)&(tk/s/W) & Impv\\
        \hline
        1.3B,1&\cite{scalable}(FPGA)  &  24& 0.05$\times$& 13.9 &2 & 0.5$\times$ \\
        \hline
        1.3B,1&A100  &  499& 1$\times$ & 119.6 &4 & 1$\times$ \\
        \hline
        1.3B,1&\textbf{{\TerEffic}(Ours)}  &  1,489& 3$\times$&  46.2& 32& $8\times$ \\
        \hline
        \hline
        1.3B,16&A100  &  7,202& 1$\times$ & 132.5 &54 & 1$\times$ \\
        \hline
        1.3B,16&\textbf{{\TerEffic} (Ours)} &  5,885& 0.8$\times$&  46.2&127& $2.4\times$ \\
        \hline
        \hline
        2.7B,1&A100  & 250&1$\times$& 124.0 & 2& 1$\times$ \\
        \hline
        2.7B,1&\textbf{{\TerEffic} (Ours)} &727&3$\times$&
        46.2&16& 8$\times$ \\
        \hline
        \hline
        2.7B,16&A100  & 3,660&1$\times$& 139.3 & 26& 1$\times$ \\
        \hline
        2.7B,16&\textbf{{\TerEffic} (Ours)} &3,028&0.8$\times$&
        46.2&66& 2.5$\times$ \\
        \hline
    \end{tabular}
    }
\end{table} 

The HBM-assisted architecture can be deemed as an ablation study of the fully on-chip design. The improvement of the fully on-chip architecture over the HBM-assisted one can be attributed to the bandwidth advantage of on-chip SRAM, while the improvement of the HBM-assisted architecture over GPUs is directly due to our specialized architecture design and hardware optimizations for ternary LLMs. 
For multi-batch tasks, where GPUs typically excel, our design continues to deliver superior performance through the batch-parallelism. While the A100 can handle a massive batch size, a batch size of 16 is sufficient for typical edge tasks, where throughput with smaller batch sizes becomes more critical.

\subsection{Comparison with the SOTA Accelerators}
While the currently biggest version of the demonstration model we adopt\cite{scalable} is 2.7B, we make a further projection on a 7B model using the same structure, with the activation dimension of 4096 and 32 layers. 7B is a commonly used LLM parameter size (e.g. LLaMA-7B\cite{llama}), and we can thereby compare our performance with the SOTA FPGA-based accelerators and commercial chips. Leveraging ternary quantization and 1.6-bit weight compression, the 7B model can fit into the 8GB HBM and therefore can be accelerated by the HBM-assisted architecture. It is capable of delivering a single-batch throughput of $\approx 290$ tokens/second with 46W power consumption, which is comparable to the typical power usage of a personal laptop. 
Our architecture shows remarkable performance advantages over the SOTA FPGA LLM accelerators. By leveraging sparsity, FlightLLM\cite{flightllm} delivers 55 tokens/second for a 7B model under 45W on the same AMD U280 FPGA, while EdgeLLM\cite{edgellm} generates 75 tokens/second for a 6B model under 51W on a bigger FPGA(AMD VCU128). Therefore, our architecture outperforms theirs by $4-5\times$ in power efficiency, taking advantage of ternary quantization.

Turning to the commercial accelerators, Apple’s newly launched M4 Max, which is specifically optimized for local AI deployment, is claimed to achieve 100 tokens/second in inference for a model of 8B parameters \cite{AppleM4}. On the other hand, some high-performance LLM accelerators, such as Groq\cite{groq} and Cerebras\cite{cerebras_1}, have also benefited from on-chip inference and achieved a remarkable throughput of several thousand tokens per second. However, these chips utilize massive sizes of on-chip SRAM (e.g., 44GB for Cerebras WSE-3 engine\cite{cerebras-2}) and consume substantial energy as high as several kilowatts, which confines their use to data centers rather than edge environments. 

In addition, the recently proposed framework bitnet.cpp\cite{bitnetcpp} has enabled efficient ternary LLM inference on CPUs. By deploying the framework, an Apple M2 CPU delivers a 15 tokens/second throughput and 1 token/second/W power efficiency for their BitNet ternary model with 7B parameters. While the BitLinear layers can be directly accelerated by {\TerEffic}'s BitLinear module, the BitNet models still require high-precision matrix multiplications in the self-attention computation(see Section \ref{background}), which leads to inefficiency in our architecture focusing on ternary matrix multiplications. 
Still, {\TerEffic} outperforms BitNet by $6 \times$ in power efficiency by comparing end-to-end inference for 7B ternary models.

\section{Conclusion}

We have introduced {\TerEffic}, an FPGA-based accelerator designed to enable efficient LLM inference by leveraging the benefits of ternary quantization. We proposed a modular architecture utilizing ternary-specific hardware optimizations to enhance efficiency. We developed two variants featuring different memory architectures: the fully on-chip architecture excels at processing smaller models, while the HBM-assisted variant extends support to larger models with strong performance. Our work establishes a foundation for future research in hardware-efficient LLM deployment, particularly in resource-constrained edge environments where power efficiency is paramount.

\clearpage

\bibliographystyle{ieeetr}
\bibliography{refs}

\end{document}